\documentclass[aps,prb,preprint,showpacs,longbibliography]{revtex4-1}
\usepackage{amssymb,amsfonts,amsmath}
\usepackage{bbm}
\usepackage{dcolumn}
\usepackage{color}
\usepackage[toc,page]{appendix}

\usepackage{graphicx}

\def\br{{\bf r}}
\def\kB{{k_{\rm B}}}
\def\rmd{{\rm d}}
\def\HF{{H_{\rm For}}}

\begin{document}

\title{First principles evaluation of phase stability in the In-Sn binary system}
\author{Michael Widom}
\affiliation{Department of Physics, Carnegie Mellon University, Pittsburgh PA 15213}

\date{\today}

\begin{abstract}
The In-Sn binary alloy system exhibits several unusual features that challenge crystallographic and thermodynamic expectations. We combine first principles total energy calculation with simple thermodynamic modeling to address two key points. First, we evaluate energies along the Bain path to interpret the discontinuous transition between the phases $\alpha$-In (Pearson type tI2) and $\beta$--In$_3$Sn (also Pearson type tI2) that are identical in symmetry. Second, we demonstrate that the solid solution phases $\beta$-In$_3$Sn and $\gamma$-InSn$_4$ (Pearson type hP1) exist at high temperatures only, and they exhibit eutectoid decompositions at low temperatures.
\end{abstract}


\maketitle

\section{Introduction}

In-Sn alloys exhibit lower melting temperatures and improved thermal fatigue as compared with Pb-Sn solders~\cite{Solder,PbFreeSolder}; they also exhibit superconductivity~\cite{InSnSC,Massalski2006}, and they form the basis for the transparent conductor indium tin oxide (ITO). In addition to their practical interest, their experimentally determined alloy phase diagram poses several scientific puzzles. This paper applies first principles total energy and band structure calculations, and simple thermodynamic modeling, to address these questions.

The assessed In-Sn binary alloy system~\cite{PD_In-Sn} exhibits four phases at room temperature and above, all with substantial composition ranges. In order of increasing fraction of Sn, the phases are: $\alpha$-In, $\beta$-In$_3$Sn (both share Pearson type tI2, ``white tin''), $\gamma$-InSn$_4$ (Pearson type hP1), and $\beta$-Sn (Pearson type tI4). There exists a different, low temperature, nonmetallic phase $\alpha$-Sn (Pearson type cF8, ``gray tin'') that has low In solubility and is stable below 286K. $\alpha$-In and $\beta$-In$_3$Sn are separated by a discontinuous transition with a narrow coexistence range around Sn fraction $x\approx 9-11\%$, despite sharing the same body-centered tetragonal structure and symmetry space group (I4/mmm). This violates the normal Landau-type model of solid-solid phase transformation that supposes group-subgroup relationsships between phases. We resolve the puzzle by evaluating the energies along the Bain path of cubic$\leftrightarrow$tetragonal deformation. We also explore the differences in interatomic bonding between the $\alpha$ and $\beta$ structures.

The experimentally reported solubility range of $\gamma$ appears nearly temperature-independent, and persists towards low temperatures. This suggests a low temperature configurational entropy, in apparent violation of the Third Law of Thermodynamics~\cite{Abriata04,Chu}. We propose that $\gamma$ actually decomposes eutectoidly as temperature drops. $\beta$ also decomposes eutectoidly. The solubility of Sn in $\alpha$-In vanishes at low temperature, while the solid solution $\beta$-Sn transforms to $\alpha$-Sn with limited In solubility.

\section{Methods}

Our calculations follow widely used methods~\cite{Chapter8}.  We utilize the Vienna Ab-Initio Simulation Package {\tt VASP}~\cite{Kresse96} to carry out first principles density functional theory (DFT) total energy calculations in the Perdew-Becke-Ernzerhof generalized gradient approximation~\cite{PBE}. We adopt projector augmented wave potentials~\cite{Blochl94,Kresse99} and maintain a fixed energy cutoff of 241.1 eV (the default for Sn).  We relax all atomic positions and lattice parameters using the {\tt PREC Accurate} precision setting, and increase our $k$-point densities until energies have converged to within 0.1 meV/atom, then carry out a final static calculation using the tetrahedron integration method. Certain other settings are discussed below as needed.

Our structures and phase diagrams are drawn from the ASM phase diagram database ~\cite{ASM} and from the Inorganic Crystal Structure Database~\cite{ICSD} (ICSD), supplemented with original publications. For solid solution phases we take 16-atom supercells at a variety of compositions and enumerate all possible configurations using enumlib~\cite{enumlib}. The supercells of the $\alpha$ and $\beta$ tI2 structures are $2\times 2\times 2$. Supercells of $\gamma$-InSn$_4$.hP1 are based on an orthorhombic supercell of the hexagonal primitive cell. For $\beta$-Sn.tI4, we take a $\sqrt{2}\times\sqrt{2}\times 2$ supercell. All configurations are fully relaxed, and only the lowest energy configuration is employed in the subsequent analysis.

Given total energies for a variety of structures, we calculate the enthalpy of formation $\Delta \HF$, which is the enthalpy of the structure relative to a tie-line connecting the ground state configurations of the pure elements~\cite{Mihal04}.  Formally, for a compound of stoichiometry In$_{1-x}$Sn$_x$ with Sn fraction $x$ we define
\begin{equation}
\label{eq:H}
\Delta \HF(x)=H({\rm In}_{1-x}{\rm Sn}_x)-((1-x) H({\rm In})+ x H({\rm Sn})),
\end{equation}
where all enthalpies are per atom.  Vertices of the convex hull of $\Delta \HF$ constitute the predicted low temperature stable structures.  For structures that lie above the convex hull, we calculate the instability energy $\Delta E$ as the enthalpy relative to the convex hull.

\section{$T\to 0$~K limit}

\begin{figure}
\centering
\includegraphics[width=4in]{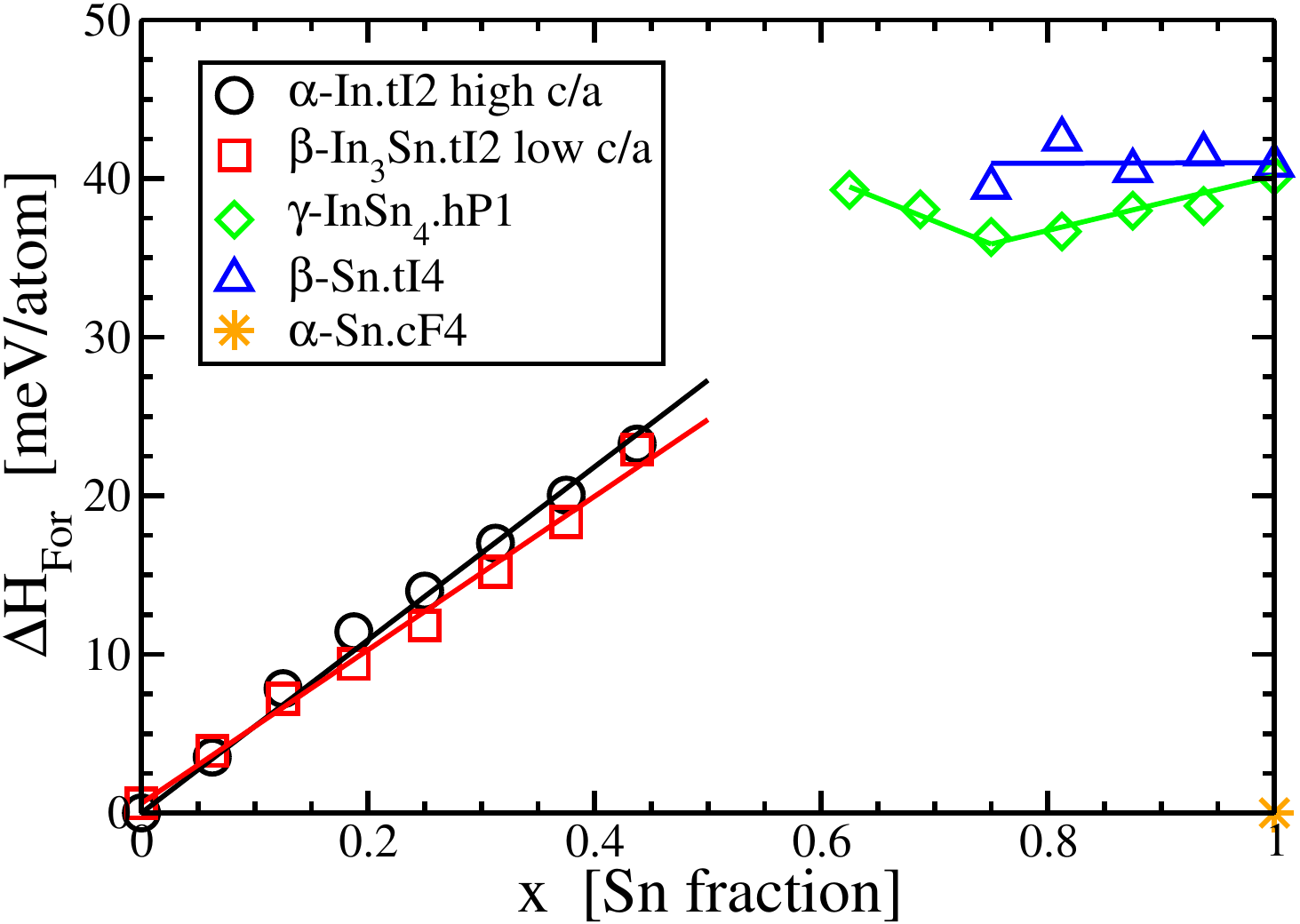}
\caption{\label{fig:nrgs}Formation enthalpies $\Delta \HF$ of In-Sn alloys calculated within density functional theory. Solid lines are linear fits to calculated enthalpies.}
\end{figure}

Composition-dependent calculated formation enthalpies are displayed in Fig.~\ref{fig:nrgs}. Notice that the known stable low temperature phases of pure In and Sn are at $\Delta H=0$, by definition, while all other formation enthalpies are positive. This implies that there are no thermodynamically stable compounds in the $T\to 0$K limit. In particular it supports the existence of a lower temperature limit for existence of the intermetallic $\beta$-In$_3$Sn phase. It also resolves the apparent third-law violation of the $\gamma$-InSn$_4$ phase by showing that it does not extend to $0$K.

Observe that the composition dependent energy of each phase is nearly linear in the concentration $x$. Small deviations from linearity reflect specific configurations representative of the solid solutions. Straight lines in Fig.~\ref{fig:nrgs} are least-squares fits constrained to pass through the endpoint at $x=0$ or $x=1$. We obtain
\begin{align}
  \Delta H_\alpha &= 0     + 0.1 x & \Delta H_\gamma &= 40 + 0.1 (x-1) \\
  \Delta H_\beta  &= 0.009 + 0.1 x & \Delta H_{\beta{\rm Sn}} &= 41 + 0.00017 (x-1) \nonumber
\end{align}
Although $\Delta H_\alpha<\Delta H_\beta$ at $x=0$, the lines cross in the vicinity of $x=0.1$ (in the middle of the experimental coexistence range), and $\beta$ is favored over $\alpha$ for larger concentrations.  The enthalpy of $\gamma$ turns up sharply for $x<3/4$ (not included in fit) because In-In neighbors cannot be avoided. The energy of $\beta$-Sn exceeds the energy of $\gamma$-InSn$_4$ for all concentrations. As $\beta$-Sn is known to be a high temperature phase, its stability must be due to some entropic effect, most likely atomic vibrations~\cite{Svib,ICQ15}.

\section{$\alpha-\beta$ transition}

\begin{figure}
\centering
\includegraphics[width=5in]{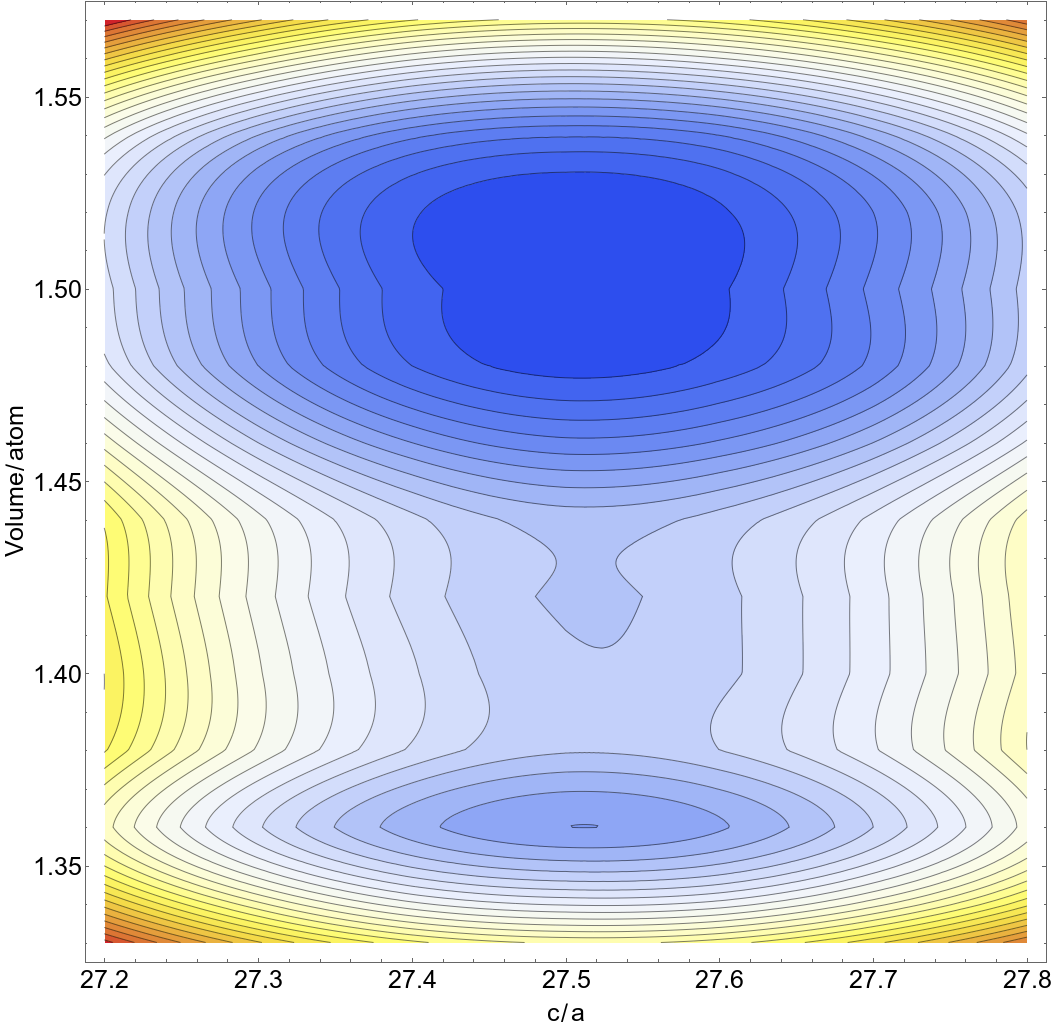}
\includegraphics[width=0.69in]{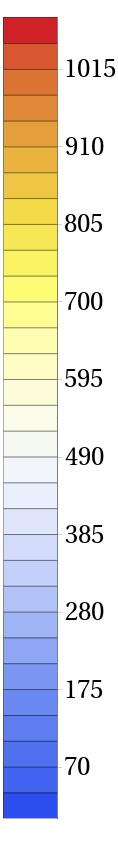}
\caption{\label{fig:sweep} Contour plot of In.tI2 energies as a function of atomic volume (horizontal axis, in \AA$^3$/atom) and $c/a$ (vertical axis).}
\end{figure}

The assessed In-Sn phase diagrams show a clear discontinuous transition from $\alpha$ to $\beta$ as the Sn fraction increases. Curiously, the phases have identical Pearson types and space group symmetries; it is unclear why they should be distinct phases. The explanation lies in their $c/a$ ratios, where $c/a>\sqrt{2}$ for $\alpha$ while $c/a<\sqrt{2}$ for $\beta$. Consider a family of tI2 structures with varying $c/a$ ratios. This is the so-called Bain path~\cite{Bain}. When $c/a=1$ the structure is body-centered cubic. At $c/a=\sqrt{2}$ it is face-centered cubic. For all other $c/a$ the structure is body-centered tetragonal.

As illustrated in Fig.~\ref{fig:sweep}, the energy of In.tI2 has a single minimum as a function of atomic volume in the vicinity of 27.5~\AA$^3$/atom, but exhibits {\em two} minima as a function of $c/a$. One deep minimum is at $c/a\approx 1.51>\sqrt{2}$, and the other is a shallow minimum at $c/a\approx 1.36<\sqrt{2}$.

\begin{figure}
\centering
\includegraphics[width=4in]{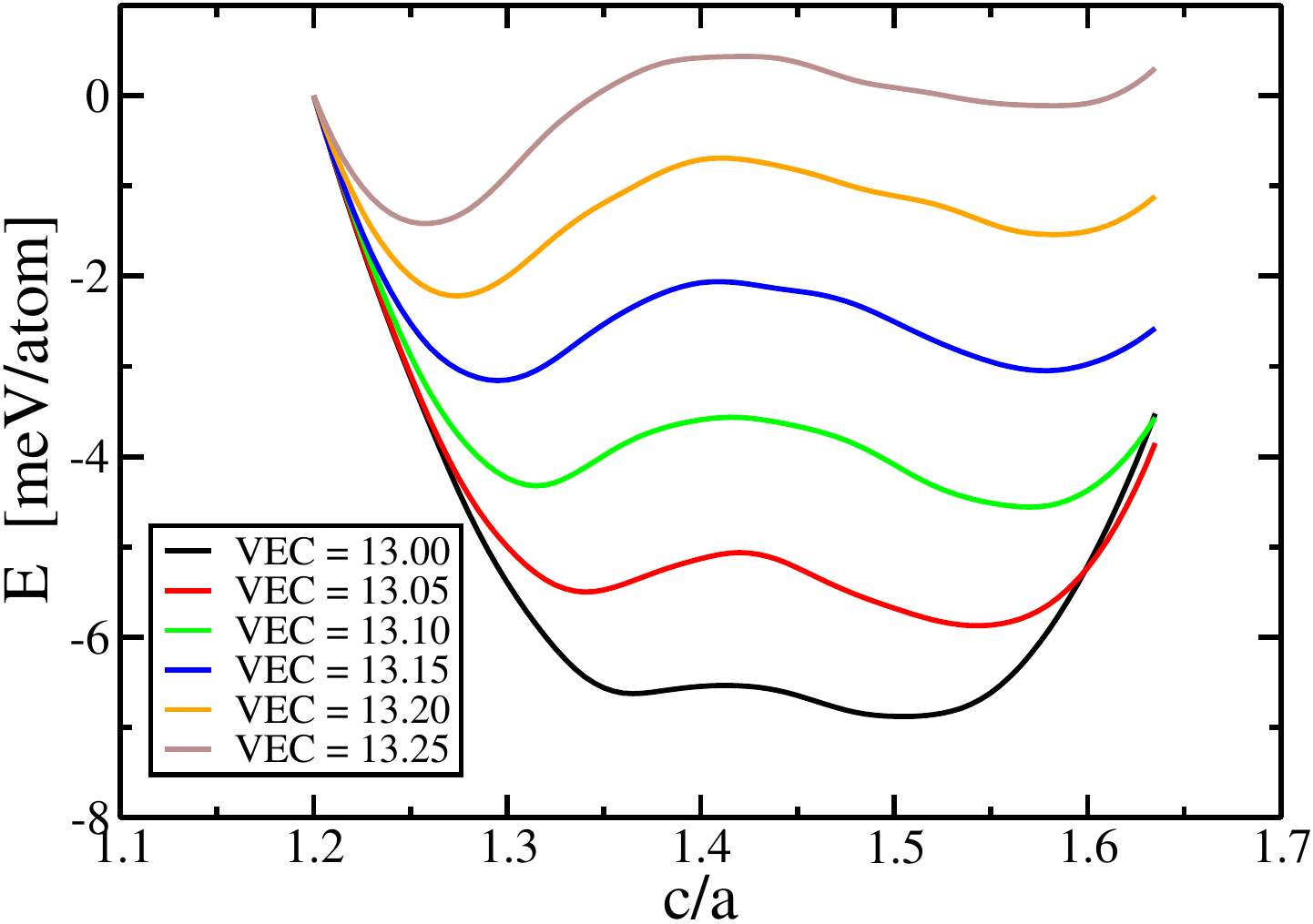}
\caption{\label{fig:Bain}Energy {\em vs.} $c/a$ ratio for tI2 structures of volume 27.5~\AA$^3$/atom.}
\end{figure}

The valence electron count (VEC) increases with $x$ because Sn lies one column to the right of In in the periodic table. We mimic the effect of alloying by increasing the electron count through the {\tt NELECT} setting of {\tt VASP}. Fig.~\ref{fig:Bain} shows how the minima evolve as functions of the VEC. For low VEC the high $c/a$ ratio is energetically prefered while high VEC prefers low $c/a$. The crossover point is around VEC$=13.125$, corresponding to concentration $x=0.125$, close to the experimental $\alpha-\beta$ coexistence range and to the crossing point in Fig.~\ref{fig:nrgs}. In all cases the two minima are separated by a barrier, which causes the $\alpha-\beta$ transition to be discontinuous.

\begin{figure*}
\includegraphics[trim=4cm 0cm 1cm 0cm,clip, width=6in]{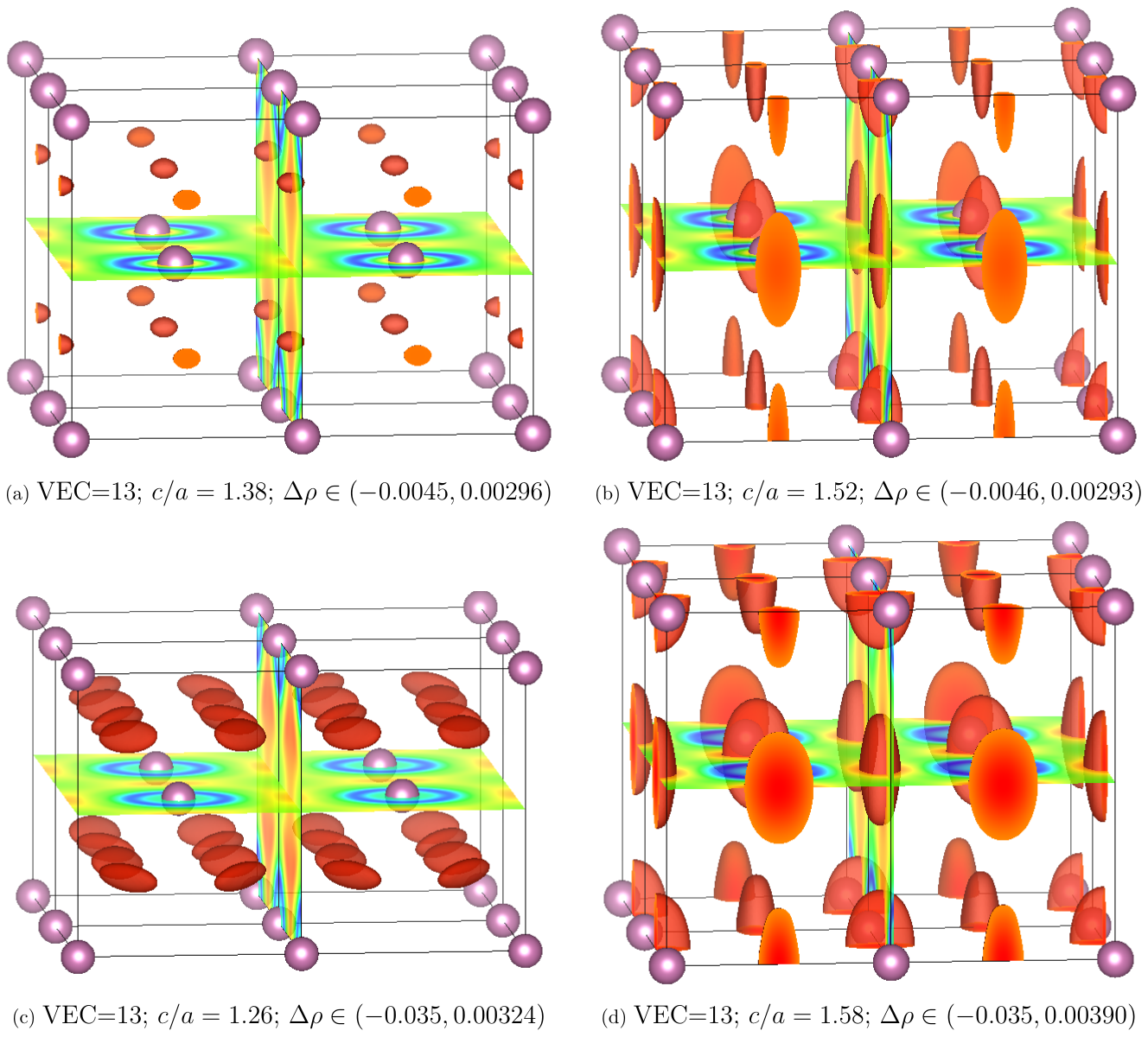}
\caption{\label{fig:charge}Difference charge densities $\Delta\rho(\br)$ at various VEC (units e/atom) and $c/a$. Cut planes and isosurfaces reveal the locus of excess charge density. Color bars range blue-green-red; isosurfaces are at 0.00195 for VEC=13 and at 0.003 for VEC=13.5.}
\end{figure*}

To explain the physical mechanism driving the $\alpha-\beta$ transition, we investigate the spatial distribution of the excess charge at elevated VEC using  the difference in charge density
\begin{equation}
  \label{eq:dQ}
  \Delta\rho(\br) = \rho_{\rm SC}(\br)-\rho_{\rm atomic}(\br).
\end{equation}
For VEC$\ne 13$, $\rho_{\rm atomic}$ must be scaled by VEC/$13$.
Fig.~\ref{fig:charge} plots $\rho(\br)$ at low/high $c/a$ with low/high VEC. All figures are at volume $27.5$~\AA$^3$/atom, and the {\tt VASP} charge density units are (electrons/\AA$^3$)$\times$(cell volume). These figures were created with the assistance of {\tt VESTA}~\cite{VESTA}. Concentrations of charge can be interpreted as covalent bonds between atoms. Evidently, at VEC$=13$/atom (the VEC of pure Indium), there is little charge accumulation at the energetically disfavored low $c/a$ (part (a)) but at the energetically preferred high $c/a$ (part (b)) charge accumulates along nearest neighbor bonds within the horizontal plane. These bonds favor a low value of $a$. In contrast, at VEC$=13.25$/atom (the VEC of In$_3$Sn) and the energetically preferred low $c/a$ (part (c)) a strong charge concentration forms along cell body diagonals. These bonds prefer to reduce the $c$ axis; the minimum bond length for fixed volume would occur at $c/a=1$.

\section{Phase diagram}

To model the composition- and temperature-dependent phase diagram we must model the set of Gibbs free energies $\{G_i(x,T)\}$ of all phases $i\in\{\alpha, \beta, \gamma, \beta-{\rm Sn},{\rm cF8\}}$, then find the convex hull of the free energies~\cite{Chapter8}. We will not concern ourselves with the liquid phase, and hence we restrict our attention to temperatures below $\approx 400$~K. Our goal is to gain qualitative understanding that reveals the physical origins of the transitions. We will treat chemical configurational (substitutional) entropy through ideal mixing,
\begin{equation}
  \label{eq:Sc}
  S(x)/\kB=-x\ln{x}-(1-x)\ln{(1-x)},
\end{equation}
and express $G(x,T)=\Delta \HF(x)-TS(x)$ for each phase $i$. As the Sn.cF8 phase is experimentally known to have low solubility of In, we restrict it to $x=1$ with $S=0$. We will also discuss, but not directly use, the vibrational free energy
\begin{equation}
  \label{eq:Sv}
  F_v=\kB T\int\rmd~\nu ~g(\nu)~ \ln[2\sinh(h\nu/2\kB T)],
\end{equation}
where $g(\nu)$ is the vibrational density of states. The electronic entropy is negligible.

\begin{figure}
\centering
\includegraphics[width=4in]{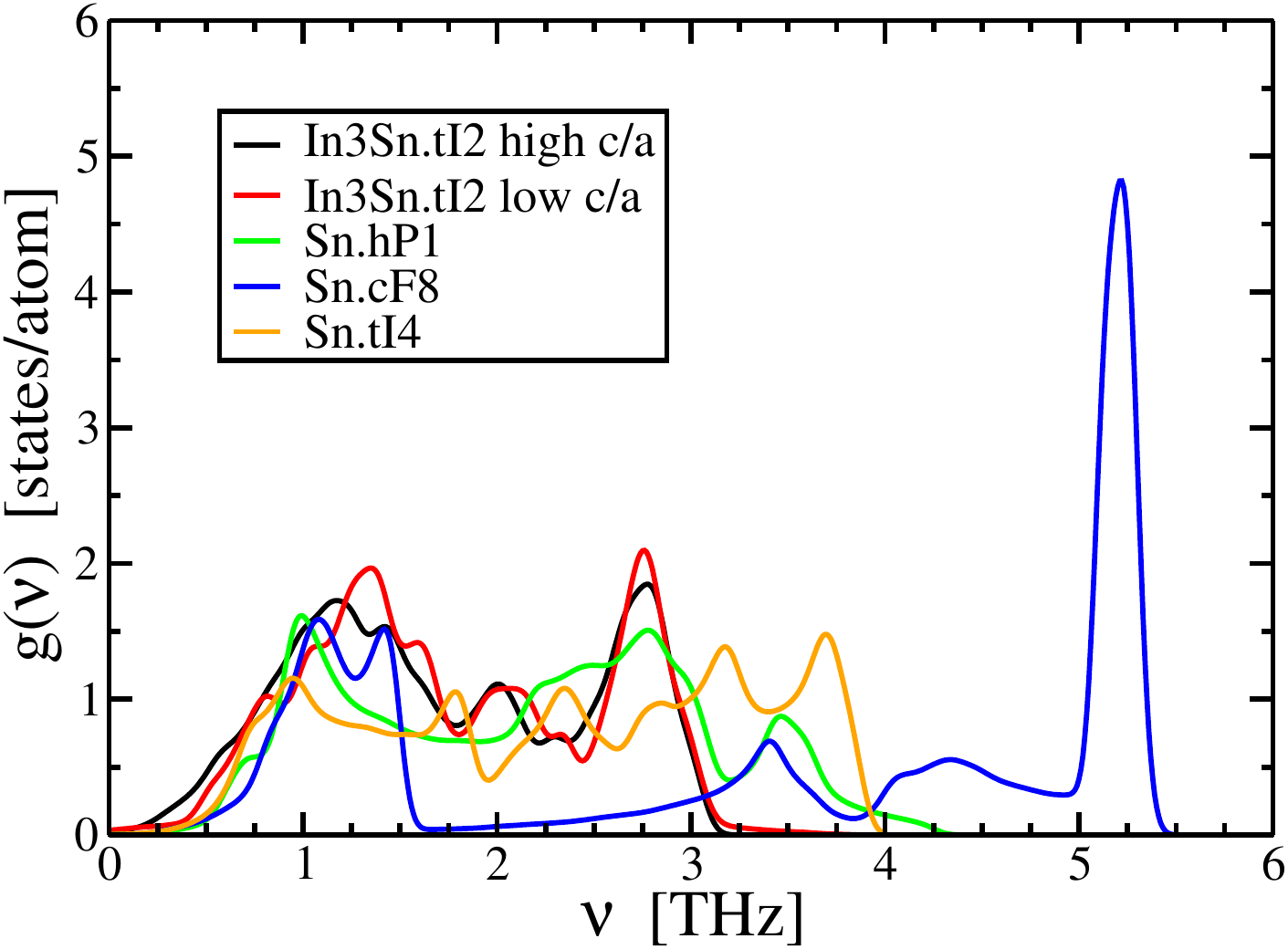}
\caption{\label{fig:vDOS} Vibrational densities of states of In$_3$Sn and elemental Sn. A Gaussian smearing of 0.05 THz has been applied for clarity.}
\end{figure}

Unfortunately, within this model, neither $\gamma$ nor the $\beta$-Sn phase is predicted to be stable within our temperature range. The difficulty can be traced to the enthalpy difference between those phases and the low temperature phase $\alpha$-Sn.cF8, which DFT predicts to be $40$ meV/atom, approximately double the experimentally reported value~\cite{Hultgren}. Alternate pseudopotentials and exchange-correlation potentials fail to resolve the matter. At $x=1$, the $\alpha$-Sn to $\beta$-Sn transition is presumably driven by vibrational entropy, and this is supported by the high frequency optical modes in the cF8 that raise the vibrational free energy (see Fig.~\ref{fig:vDOS}). Our calculated vibrational free energies place the transition at 490K, far above the experimentally reported 286K. This overestimate is not adequately resolved in the quasiharmonic approximation.

Given this difficulty we resort to ad-hoc adjustments of the energies and free energies at $x=1$. Specifically, we reduce the $x=1$ intercepts of $\gamma$ and $\beta$-Sn by 6 and 9 meV, respectively, in order to approximately match their experimental phase boundaries at 400K. Then we define a temperature-dependent free energy of cF8 to match the experimental free energy difference of Sn.cF8 and Sn.tI4~\cite{Hultgren}. Within these approximations, we identify phase boundaries by solving the equations
\begin{equation}
  \label{eq:co-exist}
  \frac{\partial G_i}{\partial x_i} =
  \frac{\partial G_j}{\partial x_j} =
  \frac{G_j(x_j)-G_i(x_i)}{x_j-x_i}.
\end{equation}
Only solutions on the convex hull are retained.

\begin{figure}
\centering
\includegraphics[width=4in]{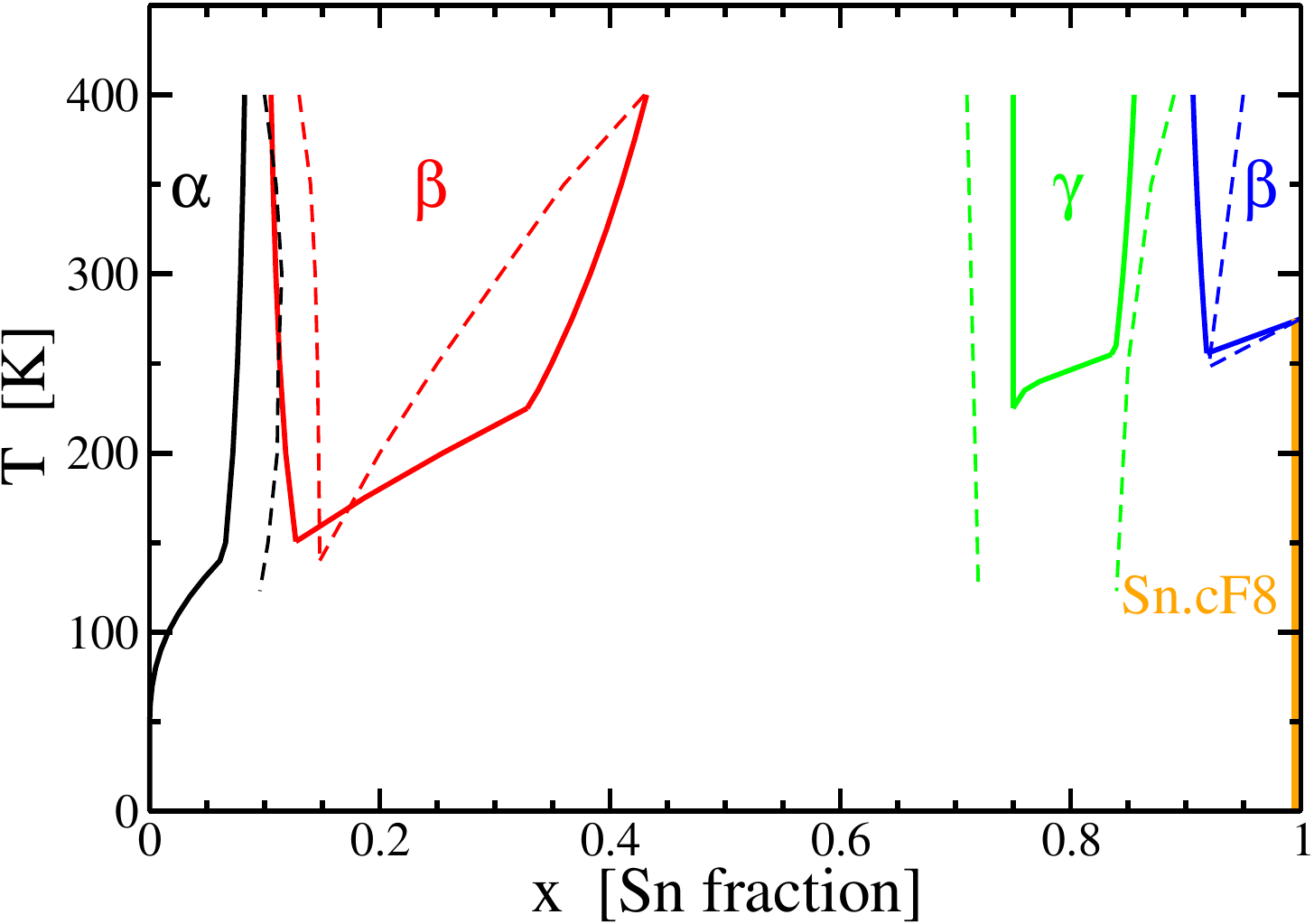}
\caption{\label{fig:phases} Predicted solid state phase diagram of In-Sn}.
\end{figure}

Figure~\ref{fig:phases} displays the resulting phase diagram. Due to the approximations made we expect only qualitative validity. The key features to observe are the low temperature eutectoid decompositions of the $\beta$ and $\gamma$ intermetallic solid solutions. These phases are stabilized by their chemical configurational entropy, and they exist at high temperatures only. Decomposition of $\beta$ has been previously suggested~\cite{PD_In-Sn}.

The stability of $\gamma$ has proven problematic. We propose it should decompose below a temperature around 245K. In contrast, most existing phase diagrams~\cite{Fisk1954,Bartram78,Evans1983,Okamoto1992} display a broad composition range extending towards low temperature. Two recent investigations~\cite{David2004InPbSn,ISOMAKI2006173}, however, do indicate a narrowing of the phase with decreasing temperature. Note that the left-hand boundary of $\gamma$ is frozen at $x=0.75$ by the kink in $\Delta \HF$. We have artificially smoothed out the approach of the right-hand boundary to this limit.

\section{Conclusion}

Our first-principles total energy calculations give insight into otherwise puzzling features of the In-Sn phase diagram. We address two main points. First, we explain the distinction between the tI2 phases, $\alpha$-In and $\beta$-In$_3$Sn on the basis of variation of the energy along the Bain path. At $c/a=\sqrt{2}$ the otherwise-tetragonal structures become face centered cubic. This symmetric structure occurs at a local maximum of the energy, with the nearby minima corresponding to $\alpha$- with $c/a>\sqrt{2}$, and $\beta$- with $c/a<\sqrt{2}$.

Second, we argue that solid solutions $\beta$-In$_3$Sn and $\gamma$-InSn$_4$ exist only at elevated temperatures, because their formation enthalpies are strictly positive. Eutectoid decomposition of $\beta$ has been demonstrated experimentally~\cite{Bartram78}, which is remarkable given the low temperature of the transformation (151K in our calculation, 140K experimentally). Eutectoid decomposition of $\gamma$ has not been conclusively demonstrated experimentally. Early phase diagrams illustrate the composition range extending towards low temperatures, although some recent studies suggest a possible narrowing of the range. We predict a nearly temperature-independent range bounded at low Sn fraction by the kink in energy at $x=3/4$, and on the right hand-side by coexistence with $\beta$-Sn. However, the range is quickly cut off below 225K due to competition with $\alpha$-Sn. $\gamma$ can transform martensitically to $\beta$-Sn~\cite{Raynor1954,Chu}, but not to $\alpha$-Sn owing to its distinct crystal structure, cF8 (diamond). Presumably the dynamics of decomposition to $\alpha$-Sn is too sluggish to readily observe.

Our goal of first-principles phase diagram prediction was not fully met, because of our inability to accurately model the $\alpha-{\rm Sn}\leftrightarrow\beta-{\rm Sn}$ transformation of pure elemental Sn. The $\gamma$-InSn$_4$ and $\beta$-Sn phases should not exist at all according to our calculated energies. Correcting this problem may require a post-DFT electronic structure methods such as the random phase approximation~\cite{VASP-RPA}. The overestimate of the cF8$\leftrightarrow$tI4 energy difference lends support to this possibility. Alternatively, fully anharmonic vibrational free energies~\cite{Svib} may prove to be required as part of the solution.

\section{Acknowledgments}
This work was supported by the Department of Energy under grant DE-SC0014506. This research also used the resources of the National Energy Research Scientific Computing Center (NERSC), a US Department of Energy Office of Science User Facility operated under contract number DE-AC02-05CH11231 using NERSC award BES-ERCAP24744.

\bibliography{refs}

\end{document}